# A Verified Algorithm Enumerating Event Structures[*]

Juliana Bowles and Marco B. Caminati

School of Computer Science, University of St Andrews
Jack Cole Building, North Haugh, St Andrews KY16 9SX, UK
{jkfb|mbc8}@st-andrews.ac.uk

**Abstract.** An event structure is a mathematical abstraction modeling concepts as causality, conflict and concurrency between events. While many other mathematical structures, including groups, topological spaces, rings, abound with algorithms and formulas to generate, enumerate and count particular sets of their members, no algorithm or formulas are known to generate or count all the possible event structures over a finite set of events. We present an algorithm to generate such a family, along with a functional implementation verified using Isabelle/HOL. As byproducts, we obtain a verified enumeration of all possible preorders and partial orders. While the integer sequences counting preorders and partial orders are already listed on OEIS (On-line Encyclopedia of Integer Sequences), the one counting event structures is not. We therefore used our algorithm to submit a formally verified addition, which has been successfully reviewed and is now part of the OEIS.

## 1 Introduction

Event structures [18, 19] model distributed systems, representing sequential-parallel behaviour, concurrency and non-determinism in a natural way. Event structures feature simple notions, namely, a set of events and binary relations over events. Amongst the relations we have *causality*, a partial order describing how the events are causally related, and *conflict* specifying which events are in conflict. In addition, a natural condition makes sure that conflict propagates over causality, i.e., events caused by conflicting events are still conflicting. Being expressed in terms of such basic concepts as partial orders and binary relations, this model has fruitful connections with combinatorics [2], topology [20], category theory [4], graph theory [2], and other causal models such as Petri nets [20].

For the same reason, basic problems regarding event structures, such as checking if a given structure is a valid event structure, whether it can be completed to a valid event structure, and event structure composition can be processed through a variety of computational tools [5, 6, 3]. However, the generation of all the possible event structures over a finite set of $N$ events is not straightforward. Currently, it is not even known how many distinct event structures there are, given a finite set of events. We will refer to these two problems as *enumeration* and *counting*, respectively. Given that the enumeration of partial orders is a hard problem, and that even the most elaborate of the known techniques for counting partial orders pass through some form of enumeration [15], the counting and enumeration problem for event structures inherit a similar amount of hardness: indeed, we are not aware of any theoretical results to count event structures which would avoid enumerating partial orders. Hence, the counting problem for the event structures motivates the enumeration problem. The naive approach consisting of generating all the possible relations and checking them for compliance to the definition of event structures is prohibitively inefficient, even for very low values of $N$. This paper presents a novel, recursive enumerating and counting algorithm which does better than that, based on an original theoretical result characterising event structures (Lemma 2); as a further contribution, the paper shows how it can be formalised and proved to be correct in the theorem prover Isabelle. The paper is organised as follows: Section 2 presents the definition of event structure, and the problem through a formal description and a simple example. Section 3 describes our algorithm, and the lemma on which it is based. Section 4 introduces the Isabelle formalisation of the involved structures, the Isabelle implementation of the algorithm, and its correctness theorems. Section 5 discusses improvements to the algorithm, and provides performance data and numerical results, while Section 6 concludes.

---

[*] This research is supported by EPSRC grant EP/M014290/1.

## 2 Description of the Problem

We recall that a pre-order on a set is a relation which is reflexive on that set and transitive, and that a partial order over a set is an antisymmetric pre-order over that set.

**Definition 1.** *An* event structure *is a triple $E = (Ev, \rightarrow^*, \#)$ where $Ev$ is a set of events and $\rightarrow^*, \# \subseteq Ev \times Ev$ are binary relations called* causality *and* conflict, *respectively. Causality $\rightarrow^*$ is a partial order over $Ev$. Conflict $\#$ is symmetric and irreflexive, and propagates over causality, i.e., $e\#e' \wedge e' \rightarrow^* e'' \Rightarrow e\#e''$ for all $e, e', e'' \in Ev$. An event $e$ may have an immediate successor $e'$ according to the order $\rightarrow^*$: in this case, we can write $e \rightarrow e'$. The relation given by $\rightarrow$ is called* immediate causality.

We will introduce a function `enumerateEs` taking one natural number $N$ as an argument, and returning a list containing, without repetitions, each possible event structure over a set of $N$ distinct events. The events are represented by the first $N$ natural numbers: $0, \ldots, N-1$. Each entry of the output list is a pair $(O, C)$ of relations over the set $\{0, \ldots, N-1\}$, the causality and the conflict, respectively. Finally, each relation will be represented by a set of ordered pairs, i.e., elements of the Cartesian product $\{0, \ldots, N-1\} \times \{0, \ldots, N-1\}$, as customarily done in set theory. This implies that, given an event structure $(E, \rightarrow^*, \#)$, and two events $e_0, e_1$, we can write $(e_0, e_1) \in \rightarrow^*$ and $(e_0, e_1) \in \#$ in lieu of $e_0 \rightarrow^* e_1$ and of $e_0 \# e_1$, respectively.

Based on `enumerateEs`, we also introduce a second function `countEs` taking the same argument, and returning a natural number counting the distinct elements of the list `enumerateEs N`.

As an example, let us illustrate the output of our algorithm for the case $N = 2$: in other words, we only have two events, 0 and 1, and we want to know all the possible valid event structures defined on these two events.

`enumerateEs 2` returns the list

```
"[({(0,0), (0,1), (1,1)}, {}), ({(0,0), (1,0), (1,1)}, {}),
  ({(0,0), (1,1)}, {(0,1), (1,0)}),({(0,0), (1,1)}, {})
 ]" :: ((nat × nat) set × (nat × nat) set) list.
```

It has four entries (what follows the character ] on the last line is not part of the list, but only specifies the type of the list itself), so, naturally, `countEs` returns 4.

Note that the first element of each pair of the list above is a partial order on $\{0, 1\}$, and hence contains at least $(0, 0)$ and $(1, 1)$ due to reflexivity. The first two pairs describe two isomorphic event structures (Figure 1, on the left): they can be obtained one from the other by swapping the numerical labels of their events. Since in this case the two events are related by causality, no conflict is possible, in light of propagation property in Definition 1: hence, the only possibility is the empty conflict relation, which corresponds to the second element of the first two pairs of the list above being the empty set.

The remaining two pairs of list above describe two event structures (Figure 1, on the right) with the same causality relation: no event is caused by the other (i.e., each event is in causal relation only with itself); in this situation, the definition of event structure allows only two possibilities. That is, either there are no conflicts, or the two events are mutually in conflict.

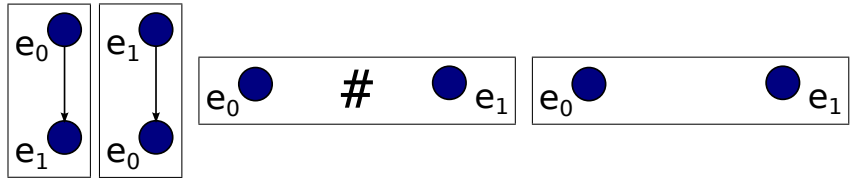

**Fig. 1.** The event structures over two events (arrows represent immediate causality)

## 3 Description of the Algorithm

The algorithm has two stages: first, all the possible partial orders on $\{0, \ldots, N-1\}$ (also called *posets*) are computed; then, for each such partial order, a list of all the allowed conflicts for it is computed. By an allowed conflict for a given partial order $o$, we mean a relation $c$ over $\{0, \ldots, N-1\}$ such that $(o, c)$ is an event structure. Once we have in place procedures to compute these two objects, obtaining our final algorithms `enumerateEs` and `countEs` is easy.

### 3.1 Computation of Posets

To enumerate all posets over $\{0, \ldots, N-1\}$, we represent them through their *adjacency matrices*. An adjacency matrix $A$ for a relation $\mathcal{R}$ over $\{0, \ldots, N-1\}$ is a square boolean matrix of dimension $N$ such that $A_{i,j} = 1$ if and only if $(i-1)\,\mathcal{R}\,(j-1)$. There is clearly a one-to-one correspondence between such matrices and all the possible relations over $\{0, \ldots, N-1\}$, therefore they can be used to represent such relations. Hence, we will say that a square boolean matrix is reflexive (respectively, transitive) when the corresponding relation is reflexive (respectively, transitive). This representation is useful in our case, because there is a simple result expressing the property that a relation is transitive and reflexive as simple conditions on the rows and columns of its adjacency matrix.

**Lemma 1.** *Assume a square, boolean matrix $B$ of order $N+1$ is partitioned as follows:*

$$B = \begin{bmatrix} A & \alpha \\ {}^\tau\!\beta & 1 \end{bmatrix},$$

*where $\alpha$ and $\beta$ are column vectors of order $N$. Then $B$ is reflexive and transitive if and only if $A$ is reflexive and transitive and, $\forall i,j \in \{1, \ldots, N\}$, 1. $(A_{i,j} \wedge \beta_i) \to \beta_j$, and 2. $(\alpha_i \wedge (\neg A_{i,j})) \to \beta_j$, and 3. $(A_{i,j} \wedge \alpha_j) \to \alpha_i$.*

We omit the straightforward proof here, and note that Lemma 1 gives a way to compute recursively all transitive and reflexive matrices, by reducing this task to computing all pairs of column vectors $(\alpha, \beta)$ satisfying the conditions (2), (1), (3). We will find that those conditions admit a computable functional translation into Isabelle/HOL, thereby allowing the recursive computation of all transitive and reflexive matrices of a given (finite) dimension $N$. Doing this corresponds to enumerating all the preorders over $\{0, \ldots, N-1\}$, which can be easily used to enumerate all the partial orders over the same set (see Section 4.2).

### 3.2 Computation of Allowed Conflicts

Let $P$ be a partial order, and denote with $F(P)$ the set of all allowed conflicts for $P$. We introduce a result permitting the recursive computation of $F$. To be able to proceed, we need some basic definitions: we denote with $P - A$ the sub-relation of $P$ obtained by removing all the pairs of $P$ whose first or second element is in the set $A$. This can be formally obtained, for example, by writing $P \backslash (A \times \operatorname{ran} P) \backslash (\operatorname{dom} P \backslash A)$, where $\backslash$ is the infix symbol for set-theoretical difference. In the following definitions, we will denote with dom and ran the domain and range of a relation; $\wp$ and $\times$ will denote the powerset operator and the infix operation symbol for Cartesian product, respectively.

**Definition 2.** $P^\to$ *is the map $A \mapsto \{y. \exists x \in A.\,(x,y) \in P\}$. That is, $P^\to(A)$ is the image of $A$ through the relation $P$. We will often omit brackets around $A$ when the latter is expressed in braces notation; e.g. $P^\to(\{a\})$ becomes $P^\to\{a\}$.*

We note that the $^\rightarrow$ operator is defined on any relation and yields a function, hence it can be iterated (e.g., $(P^\rightarrow)^\rightarrow$ always exists). Another widely used notation for $P^\rightarrow(A)$ is $P[A]$.

**Definition 3.** *Given a finite partial order $P$, $\overline{P}$ is the transitive reduction of $P$ (also known as the covering relation of $P$): $(x, z) \in \overline{P}$ if and only if $(x, z) \in P \setminus \{(x, x)\}$ and $\forall y. (x, y) \in P \land (y, z) \in P \to y = x \lor y = z$.*

In other words, the covering relation of $P$ only considers the immediate strict successors according to $P$; in yet other words, $\overline{P}$ is the smallest relation from which $P$ can be reconstructed by reflexive and transitive closure operations [1].

**Definition 4.** *$P^{-1}$ is the inverse of the relation $P$. When $P$ is a partial order, $m \in \operatorname{dom} P$ is a minimal element for $P$ if $(x, m) \in P \to x = m$.*

That is, $m$ is minimal when there are no elements smaller than $m$. We recall that for finite partial orders there is always at least one minimal element. This permits to always apply the following lemma, which is the basis for our algorithm.

**Lemma 2.** *Let $P$ be a finite partial order, and $m$ be a minimal element for $P$. Then*

$$F(P) = \bigcup_{c \in F(P - \{m\})} f(P, m, c), \tag{1}$$

*where $f(P, m, c)$ is defined as*

$$\left\{ c \cup (\{m\} \times Y) \cup (Y \times \{m\}). \right.$$

$$\left. Y \in ((P - \{m\})^\rightarrow)^\rightarrow \left( \wp \left( \operatorname{dom} P \setminus \bigcup_{M \in \overline{P}^\rightarrow \{m\}} \operatorname{dom} P \setminus c^\rightarrow \{M\} \right) \right) \right\}. \tag{2}$$

Lemma 2 is useful for our purposes because it reduces the computation of $F(P)$ to the evaluation of the function $c \mapsto f(P, m, c)$, with $c$ ranging over $F(P - \{m\})$. Since $P - \{m\}$ is still a finite partial order, and of strictly smaller cardinality than $P$ as soon as $m \in \operatorname{dom} P$ (e.g., when $m$ is minimal), we can proceed by recursion. This is also because $f$ is in turn directly computable from its three arguments: it only involves set-theoretical operations, which are straightforward to compute in HOL.

To illustrate how we will exploit the lemma, consider the simple example of an event structure presenting an event having no successors. Then, we can pick this event as the $m$ appearing in Lemma 2: the argument of $\wp$ reduces to $\operatorname{dom} P$, and we assume, since we will build a recursive algorithm, to know all the possible conflicts for the reduced causality $p := P - \{m\}$, i.e., the set $F(p)$. Lemma 2 then instructs us to consider a subset $X$ of $\operatorname{dom} P$, a possible conflict relation

$c \in F(p)$, and to extend it to $c \cup (\{m\} \times Y) \cup (Y \times \{m\})$, where $Y := p[X]$: the thesis is that by varying $X$ and $c$, we obtain all the allowed conflicts for $P$.

It should also be noted that Lemma 2 holds upon replacing $\overline{P}^{\rightarrow}\{m\}$ with any $X$ such that $\overline{P}^{\rightarrow}\{m\} \subseteq X \subseteq P^{\rightarrow}\{m\}$, but fails to hold in general if $X \subset \overline{P}^{\rightarrow}\{m\}$. Therefore, the lemma gives an optimal set on which to perform the aforementioned recursion.

*Proof (of Lemma 2).* We prove separately that $F(P) \subseteq \bigcup_{c \in F(P-\{m\})} f(P, m, c)$ and that $F(P) \supseteq \bigcup_{c \in F(P-\{m\})} f(P, m, c)$.

**Proof of** $F(P) \subseteq \bigcup_{c \in F(P-\{m\})} f(P, m, c)$**:** assume $C \in F(P)$; then, $(P, C)$ is an event structure, and therefore $(p := P - \{m\}, c := C - \{m\})$ also is, so that $c \in F(p)$. Now, consider $Y := C^{\rightarrow}\{m\}$. Given the identity $C = (C - \{m\}) \cup \{m\} \times C^{\rightarrow}\{m\} \cup (C^{-1})^{\rightarrow}\{m\} \times \{m\}$, which becomes $C = c \cup \{m\} \times Y \cup Y \times \{m\}$ due to symmetry of $C$, it suffices to show that $Y \in p^{\rightarrow \rightarrow} \left( \wp \left( \operatorname{dom} P \backslash \bigcup_{M \in \overline{P}^{\rightarrow}\{m\}} \operatorname{dom} P \backslash c^{\rightarrow} \{M\} \right) \right)$. But $P^{\rightarrow}(Y) = p^{\rightarrow}(Y) = Y$ due to the monotonicity of $C$ over $P$, hence this amounts to showing that

$$Y \subseteq \operatorname{dom} P \backslash \bigcup_{M \in \overline{P}^{\rightarrow}\{m\}} \operatorname{dom} P \backslash c^{\rightarrow} \{M\}. \tag{3}$$

When $\overline{P}^{\rightarrow}\{m\} = \emptyset$, this reduces to showing the triviality $Y \subseteq \operatorname{dom} P$. We can therefore assume $\overline{P}^{\rightarrow}\{m\} \neq \emptyset$, causing (3) to become $Y \subseteq \bigcap_{M \in \overline{P}^{\rightarrow}\{m\}} c^{\rightarrow}\{M\}$ via De Morgan rules. The latter inclusion is obvious from the propagation property of $C$.

**Proof of** $F(P) \supseteq \bigcup_{c \in F(P-\{m\})} f(P, m, c)$**:** consider $c \in F(P - \{m\})$, and $X \subseteq \operatorname{dom} P \backslash \bigcup_{M \in \overline{P}^{\rightarrow}\{m\}} \operatorname{dom} P \backslash c^{\rightarrow} \{M\}$. Set $Y := (P - \{m\})^{\rightarrow}(X)$. We need to prove that $C := c \cup \{m\} \times Y \cup Y \times \{m\}$ is a valid conflict for $P$. The symmetry of $C$ is immediate from that of $c$ and from the symmetry of its definition. The irreflexivity of $C$ follows from that of $c$ and from the fact that $m \notin Y$. It only remains to show that $C$ propagates over $P$. To this end, consider $x_1, x_2, y$ and assume that both $(x_1, x_2) \in P$ and $(x_1, y) \in C$. We want to prove that $(x_2, y) \in C$. If $x_2 = m$, then $x_1 = x_2$ and the thesis is immediate. If $x_1 = m \neq m_2$, then $y \in Y$ and $X \subseteq \bigcap_{M \in \overline{P}^{\rightarrow}\{m\}} c^{\rightarrow}\{M\}$. Moreover, there exists $M \in \overline{P}^{\rightarrow}\{m\}$ such that $(M, x_2) \in P - \{m\}$. Since $(P - \{m\}, c)$ is an event structure, $c^{\rightarrow}\{x_2\} \supseteq c^{\rightarrow}\{M\} \ni y$.

Note that proving $F(P) \subseteq \bigcup_{c \in F(P-\{m\})} f(P, m, c)$ did not require any hypothesis on $m$.

## 4  Implementation and Formalisation

The code[1] consists of Isabelle/HOL functions computing the wanted objects (preorders, posets, conflicts and event structures) and of Isabelle/HOL proofs that they compute the right objects.

---
[1] About 9.5 KSLOC, available at http://bitbucket.org/caminati/oeises.

In the case of event structures, this concretely means that we first introduce formal definitions for each side of Equation 1: the one for the left-hand side (`conflictsFor`) is close to the pen-and-paper Definition 1, while the one for the right-hand side (`conflictsFor2`) is recursive and allows us to compute the wanted results. Afterwards, we prove equality (1) itself. This separation is needed because the pen-and-paper definitions of the mathematical objects we are interested in describe them, but usually do not provide a way of constructing them [8].

In the next subsection, we give the Isabelle/HOL translations of the mathematical definitions about event structures, including `conflictsFor`; in the following ones, we introduce, respectively, the recursive construction for `conflictsFor2` and the Isabelle theorem translating equation (1), i.e., the equivalence of `conflictsFor` and `conflictsFor2`. We end the section by putting all together into the computable functions `enumerateEs` and `countEs`.

### 4.1 Formalisation of the Notion of Event Structure

The definition of `conflictsFor` is, as expected, straightforward:

```
definition "conflictsFor Or =
            {C| C. isLes Or C & Field C ⊆ Field Or}",
```

where `{C| C. ...}` is Isabelle's syntax for set comprehension. We recall that the field of a relation is the union of its domain and range (the domain, range and field of a partial order coincide, anyway). Note that we do not need to specify the set of events appearing in Definition 1 because, since we are encoding relations as set of pairs, and since the partial order is reflexive, the set of events can be always reconstructed by taking the field of it. However, since the conflict relation is not reflexive, we must impose that its field is indeed a subset of the set of events: in fact, one can find examples of valid event structures where the conflict relation field ranges from the empty set to the set of all the events.

`isLes` is a boolean predicate telling whether two relations form an event structure, and is a likewise straightforward rendition of Definition 1 in Isabelle/HOL:

```
definition "isLes Or C ⟵⟶ isMonotonicOver C Or &
              sym C & irrefl C & isPo Or",
```

where `isPo Or` is true exactly when `Or` is a partial order, and is merely a shortcut for the three conditions of reflexivity, transitivity and antisymmetry, already present (as symmetry and irreflexivity are) in Isabelle/HOL standard library. The only condition we had to specify, because not already present in Isabelle's library, is `isMonotonicOver C Or`, telling us whether the propagation condition in Definition 1 is respected; the Isabelle definition is also close to the pen-and-paper version:

```
definition "isMonotonicOver C Or =
              ∀ x y. (x,y) ∈ Or ⟶ C''{x} ⊆ C''{y}",
```

`C''{x}` being Isabelle's syntax for $C^{\rightarrow}(\{x\})$ (see Definition 2).

It is now easy to define all the event structures over a given set $X$:

```
definition "posOver X={P. Field P=X & isPo P}".
definition "esOver X={(P, C)| P C. P∈posOver X &
                                   C∈ conflictsFor P}".
```

## 4.2 Construction of Posets

We start by recursively constructing all the possible preorders over $\{0,\ldots,N-1\}$ by applying Lemma 1. This requires expressing conditions (2), (1) and (3) in that lemma into computable equivalents, which is done respectively by `cond1Comp`, `cond2Comp`, and `cond3Comp` defined as follows.

```
definition "cond1Comp matr col =
allSubLists (if (filterpositions id col = []) then
[True. i<-[0..<size (matr!0)]] else
rowAnd (map (nth matr) (filterpositions id col)))"

definition "cond2Comp matr row = ((set o concat)
 (map (filterpositions id)
      (sublist matr (set (filterpositions id row)))))
⊆ (set (filterpositions id row))"

definition "cond3Comp matr col = (set o concat)
 (map (filterpositions id)
      (sublist (List.transpose matr)
               (set (filterpositions id col))))
⊆ (set (filterpositions id col))"
```

Here, `filterpositions f l` returns a list of all the indices such that the corresponding entries of `l` satisfy `f`. `allSubLists l` returns all the possible boolean lists obtained from `l` by leaving the Falses fixed and assigning arbitrary values to the other entries. `rowAnd A` takes the logical "and" of all the rows of the matrix `A`. `sublist l I` returns the sublist of `l` obtained by taking only the entries whose indices are in the set `I`. We had to construct `filterpositions`, `allSubLists` and `rowAnd` in a computable way, while `sublist` is pre-defined in Isabelle libraries. `concat` concatenates a list of lists, while `set` converts a list to a set.

Now, we refer to Lemma 1 to define a function enumerating all preorders on $\{0,\ldots,N-1\}$, by enumerating all transitive and reflexive square matrices of order $N$:

```
fun enumeratePreorders where
"enumeratePreorders 0 = [[[]]]" |
"enumeratePreorders (Suc n) = concat (map (split auxFun)
[(matr,col). matr<-enumeratePreorders n,
  col<-filter (cond3Comp matr)
         (allSubLists [True. n<-[0..<size (matr!0)]])])",
```

where `split` uncurries its argument, and where list comprehension notation has been used.[2] `auxFun` is an auxiliary function defined as:

```
definition "auxFun matr col =
map (appendAsColumn (col@[True]))
    (map ( l. matr@[l]) (filter (cond2Comp matr)
                                (cond1Comp matr col)))",
```

and making use of `cond1Comp` and of `cond2Comp`, along with the function `appendAsColumn c A`, which appends to the matrix `A` the column `c`; `filter f l` returns the list of entries of `l` which verify `f`, while `@` is the infix operator for list concatenation. To go from the enumeration of preorders to that of partial orders, we pass through the enumeration of strict preorders and of strict partial orders as intermediate steps. First, we can enumerate strict preorders as follows:

```
definition "enumerateStrictPreorders N =
            map (setDiag False) (enumeratePreorders N)",
```

where `setDiag a A` sets to `a` all the diagonal entries of the matrix `A`. Now, we can obtain the enumeration of all strict partial orders:

```
definition "enumerateStrictPOs N = filter
  (  adjMatr. symmetricEntries adjMatr = {})
  (enumerateStrictPreorders N)",
```

where `symmetricEntries A` is a computable function returning the pairs of indices corresponding to the symmetric entries of the matrix `A` (we omit its definition here). Finally, we can enumerate all partial orders by:

```
definition "enumeratePo N =
  map (adj2PairList o (setDiag True))
      (enumerateStrictPOs N)",
```

where `adj2PairList` converts back from the adjacency matrix to the corresponding order relation, expressed as a list of pairs over the set $\{0, \ldots, N-1\}$.

### 4.3 Construction of Event Structures

The definition of `conflictsFor2`

```
function conflictsFor2 where "conflictsFor2 P=
(let (M,p)=ReducedRelation P in if M=None then [{}] else
concat [remdups (generateConflicts (set P) (the M) c).
       c<-conflictsFor2 p])"
```

reproduces the right-hand side of equation (1) in Isabelle: the union becomes `concat`, $f$ becomes `generateConflicts`, and the quantifier for the union operation is represented by `c<-conflictsFor2 p`. First, `ReducedRelation P` takes care of finding a minimal element $m$ for the partial order `P`, which is needed to build $P - \{m\}$ appearing in (1), and returns both $m$ and $P - \{m\}$ as an ordered pair. This is slightly complicated by the fact that, although we will always use it on partial order arguments, `conflictsFor2` has to always terminate, even when `P` is not necessarily a partial order. And, when `P` is not a partial order, we no longer know what to return as $m$; to work around this problem, `ReducedRelation` returns a pair $(M, p)$, where $M$ has an `option` type. This means that we can describe the cases when a minimal element for `P` is not defined by assigning to `M` the value `None`, whilst in the other cases we can extract from `M` the actual value $m$ of the minimal element using the operator `the`. Only in this latter case $P - \{m\}$ has a definite value, which gets passed (with the name `p`) to `conflictsFor2` for recursion. Otherwise, the list containing only the empty conflict relation ($[\{\}]$) is returned immediately.

---
[2] Such a notation has the form, in the simplest but typical case, [g x. x<-l, Q x], and intuitively represents the list obtained by applying the generic function `g` to the entries of a given list `l` which satisfy a condition `Q`.

`generateConflicts` is the Isabelle version of the function $f$ appearing in Lemma 2:

```
definition "generateConflicts P m c =
[c∪({m}×Y)∪(Y×{m}). Y <-
  map (Image {z∈P. fst z≠m & snd z≠m})
      (map set (sublists
                  (if P''{m}-{m}={}
                   then sorted_list_of_set (Domain P)
                   else sorted_list_of_set
                          (⋂ {c''{M}|M. M ∈ next P {m}})))))]",
```

where `next P {m}` is the set of the immediate successors of `m` and corresponds to $\overrightarrow{P}\{m\}$ appearing in expression (2). While equation (2) and `generateConflicts`'s definition have the same structure, there are some minor differences. First, `generateConflicts` returns a list of relations, while $f$ returns a set of relations. In general, it is easier to prove theorems about sets than about lists but, on the other hand, it is easier and more efficient to compute using lists, rather than sets. For the same reason, the argument of `conflictsFor2` is a list of pairs representing a relation, while the argument of $F$ appearing in (1) is a set of pairs. Since lists contain more information than the corresponding finite sets, this is not a problem. The Isabelle function `set` goes from a list to a set, while `sorted_list_of_set` goes the other way around. When passing from a set representation of relation to a list-based one, the $\rightarrow$ operator (Definition 2) becomes the `map` operator, the powerset operator $\wp$ becomes `sublists`. Another difference with expression (2) is that it differentiates two cases for computational reasons: when $P``\{m\}-\{m\}\neq\{\}$ the argument of $\wp$ occurring in (2) can be simplified to the intersection appearing after `sorted_list_of_set`. In Isabelle, `P''A` is the infix notation for `Image P A`, where `Image` is the operator $\rightarrow$ (Definition 2). Now we can construct all the event structures over `N` elements:

```
definition "enumerateEs N=
concat (map (λ P. (map (λx. (set P, x)) (conflictsFor2 P)))
            (enumeratePo N))",
```

where `enumeratePo` enumerates all the partial orders over $\{0, \ldots, N-1\}$.

Counting all the event structures over `N` is also immediate:

**Listing 1.1.** Definition of `countEs`

```
definition "countEs N=listsum (map (size o conflictsFor2)
                                   (enumeratePo N))".
```

Note that we chose a combination of `listsum` and `size` instead of directly applying `size` to a suitable concatenation of list, due to better performance. However, this slightly complicates the proofs of the results we will see in Section 4.4.

### 4.4 Correctness Proof in Isabelle

Our way of formally verifying the correctness of our algorithm is to prove Isabelle theorems stating the equivalence between `conflictsFor` and `conflictsFor2`, between `esOver` and `enumerateEs`, between `card o esOver` and `countEs`. This spawns a series of intermediate lemmas to be proven, stating that `enumerateStrictPOs`, `enumeratePo`, `enumeratePreorders` all return the expected entities. Additionally, all the functions converting from one representation to another (e.g., from a list of pairs to the corresponding adjacency matrix, and the other way around) need to be proven correct. Here, we describe the milestones in this chain of proofs.

To prove the correctness of the function `enumeratePreorders` (Section 4.2), we formally proved Lemma 1, and then formally proved the correspondence between `cond1Comp` and condition (2), between `cond2Comp` and condition (1), and between `cond3Comp` and condition (3). This allowed us to prove the following theorem:

```
theorem correctnessThm: assumes "N>0" shows
"set (enumeratePreorders N) = {a| a. N=size a &
  isRectangular N a & transMatr a N & reflMatr a N}",
```

which says that the set of matrices obtained via the computable function `enumeratePreorders` matches the set of all transitive and reflexive matrices. Note that, since we are encoding a matrix in Isabelle as a list of rows, we have to make sure that each row has the same size. This is the goal of `isRectangular`. The theorem above also has a version expressed in the language of list of pairs, without using matrices:

```
theorem "set (map (set o adj2PairList) (enumeratePreorders N))
       = {r| r. trans r & refl_on {0..<N} r}".
```

The correctness theorem for partial orders generation is:

```
theorem poCorrectness: assumes "N>0" shows
"set (map set (enumeratePo N)) = posOver {0..<N}".
```

The theorem above is the first component of the proof of the correctness for `enumerateEs`. The other component is the correctness proof for `conflictsFor2`; this is attained by formally proving Lemma 2 and then applying it to the recursive definition of `conflictsFor2`. The result is:

```
theorem conflictCorrectness: assumes "trans (set Or)"
"reflex (set Or)" "antisym (set Or)" shows
"set (conflictsFor2 Or)=conflictsFor (set Or)".
```

Putting together `conflictCorrectness` and `poCorrectness`, we obtain

```
theorem esCorrectness: assumes "N>(0::nat)" shows
                      "set (enumerateEs N)= (esOver {0..<N})".
```

## 5 Performance and Results

Since we enumerate, and not only count, all the event structures, we face a fundamental complexity problem: the enumeration of event structures entails that of posets, and the latter has exponential complexity [13].

However, some considerations led us to attain for our algorithm some performance improvements which, although marginal in the general case, turned out to be significant for the computations related to the small values of $N$ that we were able to obtain. Originally, the definition for `conflictsFor2` was more naive than the one we introduced in Section 4.1:

```
function conflictsFor2v0 where "conflictsFor2v0 Or=
 (let (M,or)=ReducedRelation Or in if M=None then [{}]
  else let m=the M in concat [(generateConflicts (set Or) m c).
                              c<-conflictsFor2v0 or])".
```

The key difference is that we did not take care of removing duplicate entries in the recursive generation of the wanted event structures. At a purely mathematical level, this does not matter much since we eventually convert the obtained list to a set: see, e.g., the statement of `esCorrectness`, Section 4.4. From an algorithmic point of view, however, this does matter, since keeping duplicate entries represents a computational burden which gets amplified through recursion. It should also be noticed that, with the definition above, the definition of `countEs` needed to be different, since we do not want to count duplicates, and therefore was based on a `card o set` operation, rather than on a combination of `listsum` and of `size`, as in Listing 1.1. The problem coming from duplicates is typical of formal enumerations and counting algorithms [12]. Therefore, a second attempt was tried:

```
function conflictsFor2v1 where "conflictsFor2v1 Or=
 (let (M,or)=ReducedRelation Or in if M=None then [{}]
  else let m=the M in (remdups o concat)
    [(generateConflicts (set Or) m c).
      c <- conflictsFor2v1 or])",
```

whereby a `remdups` operation was simply added at the end of each recursion to remove duplicate entries. This measure improved the performance of our algorithm; moreover, it eliminated the need of resorting to a `card o set` application when calculating `countEs`. This is because `remdups` guarantees that there are no repetitions, and hence, to count the number of event structures, we can directly count the number of elements of the list returned by `countEs`. However, since `remdups` has quadratic time complexity [14], it pays off to invert the order of `remdups` and `concat`, which leads to the final definition of `conflictsFor2`. It should be noted that implementing this optimisation comes at a price: since it is no longer obvious that the output of `conflictsFor2` has no repetitions, we need to formally prove this; once we do that, we can prove correctness for `countEs`:

```
theorem assumes "N>0" shows "card(esOver{0..<N})=countEs N".
```

Orthogonal to the optimisation coming from a thoughtful placement of `remdups`, there is another one, which we proceed to explain. Lemma 2 holds for any $m$ chosen among the minimal elements of the partial order $P$; this reflects in the definition of `conflictsFor2`, where `ReducedRelation` performs this choice to compute the recursion step. The selected minimal element, `the M`, is passed to `generateConflicts`, where it is used as a pivot to generate the new conflict relations. While this generation is always correct as long as `the M` is indeed a minimal element, the efficiency of the corresponding computation does vary according the the choice of the pivot. Indeed, there is a noticeable difference in performance between choosing the pivot without a criterion and with a reasonable one, as we proceed to illustrate. The final `ReducedRelation` does use a criterion:

```
definition "ReducedRelation P=(let minRel=
 List.filter (λ(x,y). x ∉ (snd'((set P)−{(x,x)}))) P in
 let Minimals=remdups (map fst minRel) in
 let Multiplicities=
  map (λz. (size(Next P [z]),z)) Minimals in
 let suitableVals=Maxs(set (map fst Multiplicities)) in
 let somePivot=optApp snd (List.find (λ(x,y).x∈suitableVals)
                                     Multiplicities) in
 (somePivot, filter
   (λ(x,y). x ≠ the somePivot & y ≠ the somePivot) P))".
```

`Next P [z]` computes $\overrightarrow{P}\{z\}$, so that `Multiplicities` contains each minimal element of `P` along with the number of its immediate successors. This is used to select (through the function `Maxs`) one minimal element having the highest number of immediate successors (`optApp` and `List.find` break possible ties in case such an element is not unique). Finally, the picked value $m$ (in the form of `the somePivot`) and the reduced relation $P - \{m\}$ are returned as a pair. The idea here is that, since in the definition of `generateConflicts` there is an intersection over all the immediate successors of the pivot, by maximising the number of successors, the resulting intersection will tend to be smaller. And this is desirable, because an expensive operation of powerset will be taken on the resulting intersection. On the other hand, the original implementation of `ReducedRelation` was the most naive possible with no particular criterion for the selection of the pivot:

```
definition "ReducedRelationV0 P=(let min=
 (List.find (λx. x ∉ (snd'((set P)−{(x,x)}))) (map fst P))
 in (min, filter (λ(x,y). x ≠ the min & y ≠ the min) P))"
```

Table 1 compares how the execution time of `countEs` 5 changes when adopting `conflictsFor2v0`, `conflictsFor2v1` and the final `conflictsFor`. Table 2 compares how the execution time of `countEs` 6 (using `conflictsFor`) changes when adopting `ReducedRelationV0` and `ReducedRelation`. Times were averaged on three runs, after discarding the first one, and were taken on the same machine (dual core Intel Core 2 @ 2.40GHz).

**Table 1.** Seconds to compute `countEs` 5

| conflictsFor2v0 | conflictsFor2v1 | conflictsFor2 |
|---|---|---|
| 222 | 10 | 2 |

**Table 2.** Seconds to compute `countEs` 6

| ReducedRelationV0 | ReducedRelation |
|---|---|
| 94 | 79 |

Finally, we present in table 5 the results (for $N \in \{0, \ldots, 7\}$) of counting all the preorders, partial orders and event structures through the verified algorithms we presented. `countEs` 7 took about 55 hours to compute.

**Table 3.** Counting results

| N | preorders | partial orders | event structures | N | preorders | partial orders | event structures |
|---|---|---|---|---|---|---|---|
| 0 | 1 | 1 | 1 | 4 | 355 | 219 | 916 |
| 1 | 1 | 1 | 1 | 5 | 6942 | 4231 | 41099 |
| 2 | 4 | 3 | 4 | 6 | 209527 | 130023 | 3528258 |
| 3 | 29 | 19 | 41 | 7 | 9535241 | 6129859 | 561658287 |

## 6 Conclusions

We provided an original algorithm to enumerate and count event structures, and formally proved its correctness along with the correctness of its implementation. An immediate application of our work was the realisation that the integer sequence counting the event structures is not listed on the OEIS [16], and has no trivial or immediate link to known sequences; this was confirmed by OEIS' *Superseeker* service, which "will try hard to find an explanation for your sequence" [16]. We therefore submitted it to the OEIS; it passed the review process and is now published at http://oeis.org/A284276. Although there is existing work on the verified enumeration and counting of mathematical objects [11, 7, 12], this is the first mechanically certified addition to the OEIS we are aware of. Another immediate consequence upon looking at the obtained sequence is that even the counting problem is likely to be as hard as that of counting posets, given the fact that no advanced result permitting to detach the numbering of event structures from that of posets is known, and given the fast growth of the sequence, testified by the last column of table 5. In the process, we also supplied formal correctness theorems for two existing OEIS sequences (`A000798` and `A001035`).

Furthermore, we not only provided counting, but full enumeration of preorders, posets and event structures; we believe that this can be of interest for further investigations, in a way similar to how the enumeration of structures can be exploited for their classification and further investigations [10, 17], as well as for simulations [9].